\documentstyle[12pt,aaspp4]{article}

\begin{document}
\bigskip

\title{
New Nuclear Reaction Flow during r-Process Nucleosynthesis \\
in Supernovae:\\
Critical Role of Light Neutron-Rich Nuclei
}
\author{M.\ Terasawa$^{a-c}$, K.\ Sumiyoshi$^{c,d}$, T.\ Kajino$^{a,b}$, \\
 G.\ J.\ Mathews$^{e}$, and I.\ Tanihata$^{c}$}

\begin{center}
$^{a}$Department of Astronomy, School of Science, University of Tokyo, \\
Hongo, Bunkyo-ku, Tokyo 113-0033, Japan \\
$^{b}$National Astronomical Observatory, and \\
Graduate University for Advanced Studies, \\
Osawa, Mitaka, Tokyo 181-8588, Japan \\
$^{c}$Institute of Physical and Chemical Research (RIKEN), \\
Hirosawa, Wako, Saitama 351-0198, Japan \\
$^{d}$Numazu College of Technology, \\
Ooka, Numazu, Shizuoka 410-8501, Japan \\
$^{e}$Department of Physics and Center for Astrophysics, University of
Notre Dame, \\
Notre Dame, IN 46556, U.S.A.
\end{center}

\begin{abstract}
  We study the role of light neutron-rich nuclei during r-process
  nucleosynthesis in supernovae. Most previous studies of the r-process
  have concentrated on the reaction flow of heavy unstable nuclei.
  Although the nuclear reaction network includes a few thousand heavy
  nuclei, only limited reaction flow through light-mass nuclei near the
  stability line has been used in those studies.  However, in a viable
  scenario of the r-process in neutrino-driven winds, the initial
  condition is a high-entropy hot plasma consisting of neutrons,
  protons, and electron-positron pairs experiencing an intense flux of
  neutrinos. In such environments light-mass nuclei as well as heavy
  nuclei are expected to play important roles in the production of seed
  nuclei and r-process elements.  Thus, we have extended our fully
  implicit nuclear reaction network so that it includes all nuclei up to
  the neutron drip line for Z $ \leq 10$, in addition to a larger network
  for Z $ \geq 10$.  In the present nucleosynthesis study, we utilize a
  wind model of massive SNeII explosions to study the effects of this
  extended network.  We find that a new nuclear-reaction flow path opens
  in the very light neutron-rich region.  This new nuclear reaction flow
  can change the final heavy-element abundances by as much as an order
  of magnitude.
\end{abstract}

\newpage

\section{Introduction}
The r-process is responsible for
roughly half of the abundance of nuclei heavier than iron.  However, the
astrophysical site for this nucleosynthesis process is still a mystery
which remains as a major focus of nuclear astrophysics.

Recent detections of the r-process elements in several metal-deficient
halo stars (Sneden et al. 1996, 1998, 2000) have indicated that the
observed abundance pattern of heavy elements is very similar to that of
the solar r-process abundance (K$\ddot{\rm a}$ppeler et al. 1989,
Arlandini et al. 1999) for the mass region $120 \leq$ A. This finding
suggests that the r-process occurs in a specific environment such that the
abundance pattern is completely independent of the metallicity of the
progenitor stars. It is generally believed that the r-process occurs
under explosive conditions at high neutron density and high
temperature. It has been discussed, for sometime, that core-collapse
supernovae (type II or type Ib) could provide the most likely
environment for such r-process nucleosynthesis.  In a supernova
explosion, it is now commonly accepted that massive Fe cores do not readily
explode in a purely hydrodynamical way, but that they require
help from neutrino heating [the so-called delayed explosion (Bethe $\&$
Wilson 1985)]. The r-process occurs in the region between the surface of
the newly-formed neutron star and the outward moving shock wave (Meyer et
al. 1992). In this region, the entropy is so high that the NSE favors
abundant free neutrons and alphas rather than heavy nuclei. This is an
ideal site in that it naturally satisfies the observed metallicity
independence of the r-process yields.

Woosley et al. (1994) have performed an r-process simulation based on a
delayed explosion model, from which an excellent fit to the solar
r-process abundance pattern was obtained.  However, the required high
entropy in their supernova simulation has not been duplicated by other
numerical calculations (Witti et al. 1994, Takahashi et
al. 1994). Furthermore all those calculations used a limited network for
light nuclei and they did not consider neutrino interactions.  Since
neutrinos can completely dominate the environment just outside a newly
born neutron star, their effects must be included in nucleosynthesis
calculations. Neutrino-nucleus interaction processes during the
r-process have been considered by several authors (Meyer et
al. 1992, Meyer 1995, Fuller $\&$ Meyer 1995, Qian et al. 1997,
McLaughlin et al. 1996).  These studies have shown that, among other
things, neutrino processes tend to hinder the r-process by decreasing
the neutron-to-seed abundance ratio although they can help to smooth the
final abundance pattern.

These results may restrict the supernova explosion model. However,
Cardall $\&$ Fuller (1997), Qian $\&$ Woosley (1996), Otsuki et al.
(1999), and Sumiyoshi et al. (2000) have shown that a short
dynamic-time-scale model plus general relativistic effects can lead to a
successful r-process. This is because the temperature and density
decrease very fast. Therefore, charged-particle reactions to make seed
nuclei do not proceed efficiently, and only a small amount of seed
nuclei are produced.  Thus, the neutron-to-seed abundance ratio becomes
large enough for heavy r-process nuclei to be synthesized.

The seed nuclei in the neutrino-driven wind are produced early in the
expansion by alpha-capture processes. When the temperature and density
become low and charged-particle reactions almost cease, the r-process
starts from these seed nuclei. This is believed to be a general scenario
for r-process nucleosynthesis. Thus, in most previous studies of the
r-process, interest was paid mainly to heavy nuclei. Although a few
thousand heavy nuclei were included in the nuclear reaction network,
only a limited number of light-mass nuclei were selected near the
$\beta$-stability line.  However, the success of wind-models with a
short dynamic time scale requires that attention be given to the
reactions of light neutron-rich nuclei in a very neutron rich
environment.  Light-mass nuclei as well as heavy nuclei are expected to
play important roles in the production of seed nuclei and r-process
elements.  Indeed, it has been noted (Cameron 2001)
that alternate reaction flow paths involving neutron-rich light nuclei
may be important for $r$-process nucleosynthesis.

In order to study quantitatively the role of light neutron-rich nuclei,
we have therefore extended the nuclear reaction network. We have added about 40
unstable nuclei for Z $\leq 10$ to a larger network for Z $ \ge 10$. We
find that a new nuclear reaction paths open in the very light
neutron-rich region. We also find that these new nuclear reaction paths
can change the heavy element abundances by as much as an order of
magnitude, while still keeping the prominent three peaks of the
r-process elements as well as the hill of the rare-earth elements.  In
the present nucleosynthesis study, we have analyzed the effect of this
expanded network in the framework of a numerical simulation of the
neutrino-driven wind.

\section{Reaction Network}

The nuclear reaction network used in Meyer et al. (1992) and Woosley et
al. (1994) is probably adequate for simulating the nucleosynthesis of
intermediate-to-heavy mass nuclei. However, in the mass region Z$ \le
10$, this network is limited to only a few neutron-rich unstable nuclei
in addition to the stable ones (see Table 1).  Charged-particle
reactions, which assemble $\alpha$-particles into heavier nuclei (i.e.
$\alpha$-process), are fast at high temperature $2 \leq$ T$_9$ in the
early stage of the expanding neutrino-driven winds.  Therefore, the
following reactions (and their inverse) linking the light elements up to
$^{20}$Ne were identified to be most important: $\alpha(\alpha
n,\gamma)^9\rm{Be}(\alpha,n)^{12}\rm{C}$,
$\alpha(\alpha\alpha,\gamma)^{12}\rm{C}$,
$^{12}\rm{C}(n,\gamma)^{13}\rm{C}$,
$^{12}\rm{C}(\alpha,\gamma)^{16}\rm{O}$,
$^{13}\rm{C}(\alpha,n)^{16}\rm{O}$, $^{16}\rm{O}(n,\gamma) ^{17}\rm{O}$,
$^{16}\rm{O}(\alpha,\gamma)^{20}\rm{Ne}$, $^{17}\rm{O} (\alpha,n)
^{20}\rm{Ne}$.  On the other hand, the onset of the r-process is thought
to be delayed until the temperature drops to below T$_9 \approx 2$. By
this time many seed nuclei in the range of $70 \leq \rm A \leq 120$ have
been produced by charged-particle reactions. This is the reason why the
light-mass neutron-rich nuclei were presumed to be unimportant in the
$\alpha$-process as well as the r-process.  However, as we will discuss
later in more detail, light nuclei can be important in the extremely
neutron-rich environment of neutrino-driven winds (Otsuki et al. 2000,
Sumiyoshi et al. 2000, Kajino et al. 2001)  where they 
play a significant roles in the production of both seed and r-process
nuclei.

We have therefore extended and improved the reaction network so that it
covers all radioactive nuclei up to the neutron-drip line for $Z < 10$,
as shown in Figure 1.  Although information is limited on the rates for
$(2n,\gamma)$ reactions, we did consider nuclei which are unbound
after an (n,$\gamma$) reaction, i.e. $^{6}$He, $^{8}$He, $^{11}$Li,
$^{14}$Be, $^{17,19}$B, $^{22}$C, etc.  Our extended network thus
includes more than 63 nuclides for $Z < 10$ and more than 200 reactions among
them, while the network used in Woosley et al. (1994) includes only 27
nuclides, most of which are stable.  We included all charged-particle
reactions for A$\leq 28$, in order to study both the
$\alpha$-process and the neutron-capture flow, as well as their
competition in the production of seed nuclei.  We take the rates of
charged-particle reactions from those of Kajino et al. (1990ab), Orito
et al. (1997), and the NACRE compilation (Angulo et al. 1999). The
$\beta$-decay lifetimes are from Tachibana et al. (1990, 1995).  We also
added many heavier, neutron-rich nuclei for Z $ > 10$ from
$\beta$-stability to the neutron-drip line in addition to our extended
network code.  The total number of nuclides up to Americium isotopes is
$3036$.  We refer to this hereafter as the "full network".  We believe
we have included all possible relevant reactions in this network.

We also used another smaller network which is similar to the ones used
in Meyer et al. (1992) and Woosley et al. (1994). This network includes
only a few light neutron-rich nuclei.  We shall refer to this as the
"small $\alpha$-network".  Details on the difference between the "full
network" and the "small $\alpha$-network" are shown in Table 1 and
Figure 1.

Except for the above modifications and extension of our network, the
calculation is essentially the same as that of Meyer et al. (1992)
and Woosley et al. (1994).  Our "full network" includes ($\alpha$,n)
reactions and their inverse up to Z $=36$.  Neutron captures, their
inverse reactions, and $\beta$-decays are included for all isotopes.
Rates for these reactions are taken from Caughlan \& Fowler (1988),
Woosley et al. (1978), and OAP-422 (Woosley et al. 1975).  Neutron
capture rates for the heavier nuclei were taken from experiment where
known, and otherwise are from Holmes et al. (1976) and Woosley et
al. (1978).  The $\beta$-decay rates were taken from
Klapdor et al. (1984). We include $\beta$-delayed neutron
emission of up to three neutrons (Thielemann et al. 1983).  We use the
nuclear-mass table from Hilf et al. (1976).

As for neutrino interactions, we include electron-type neutrino capture
($\nu_e + (Z,A) \rightarrow (Z+1,A) + e^-$) for all nuclei (Qian et
al. 1997), and free neutrons ($\nu_e + n \rightarrow p + e^-$), and
electron-type antineutrino capture ($\bar{\nu}_e + p \rightarrow n +
e^+$) for free protons.  These latter two neutrino interactions
predominantly control the electron fraction, Y$_e$, during r-process
nucleosynthesis.  Neutron emission after neutrino-induced excitations
can occur.  For very neutron-rich nuclei, up to several 
neutrons can be emitted.  We also included these
processes following the method of Meyer et al. (1998).

It is noteworthy that the previous r-process calculations of Meyer et
al. (1992), and Woosley et al. (1994) had the additional complexity that
the seed abundance distribution was first calculated by using a smaller
network for light-to-intermediate mass elements, and then the result was
connected to another r-process network in a different set of
calculations. This separation was imposed because it was thought to be
numerically more efficient to run the $\alpha$-process and the r-process
separately.  However it was perhaps more difficult to interpret the
whole nucleosynthesis process.  Our nucleosynthesis calculation is
completely free from this complexity. We have exploited a fully implicit
single network code which is applied to the whole sequence from NSE to
the $\alpha$-process to the r-process.

\section{Neutrino-Driven Wind Model}

\subsection{Hydrodynamic Simulation}

Our present purpose is to illustrate the differences between
calculations in our extended network and those of
the generally employed smaller network.  For purposes of this
illustration the details of the wind model employed are
not particularly important.  We choose a model, however,
which is both derived from a ''realistic`` hydrodynamic
simulation and one which exemplifies the possible effects.

As a model for the expanding material, we employ results from the
numerical simulation of the neutrino-driven winds of Sumiyoshi et
al. (2000).  After the supernova core bounce, the proto-neutron star
emits an intense flux of neutrinos during a Kelvin-Helmholtz cooling
phase.  Some of those neutrinos heat the surface material of the
proto-neutron star.  The surface is gradually ejected from the neutron
star, forming a neutrino-driven wind. Qian and Woosley (1996) and Otsuki
et al. (2000) have studied such winds above the proto-neutron star by
solving the steady-state hydrodynamical equations including neutrino
heating and cooling. Otsuki et al. (2000) have included a
general-relativistic treatment. They obtained the time evolution of the
ejected material for a series of different neutrino luminosities and
proto-neutron-star models.  They deduced that the wind models with a short
dynamic time scale lead to successful r-process nucleosynthesis even for
an entropy of S/k$_{B}$ $\sim 140$.  This is less than that required
by Woosley et al. (1994). Sumiyoshi et al. (2000)
have confirmed this finding in their fully general-relativistic
hydrodynamical simulations without the approximation of steady-state flow
for the neutrino-driven wind.

The adopted wind model in the present illustration will consist
of a single trajectory which produces significant heavy-element
abundances.  It thus, has a very short
expansion time scale, $\tau_{dyn}=5.1 \times 10^{-3}$ s, because of the
intense neutrino flux assumed and general relativistic effects. The average
energy of electron-type neutrinos is set equal to $10$ MeV. For
electron-type antineutrinos it is $20$ MeV, and for $\mu$- and
$\tau$-neutrinos and anti-neutrinos it is $30$ MeV.  This 
is the same as has been adopted in 
previous simulations (Qian and Woosley 1996; Otsuki et al. 2000).
The total neutrino luminosity is taken to be $6 \times 10^{52}$ erg
s$^{-1}$. 

 Regarding our adopted  time scale,
it has been proposed, e.g. Meyer \& Brown (1997)
that for a sufficiently fast time scale in the wind, the neutrons
and protons may not completely reassemble to form alpha particles.  
The $r$-process might then be facilitated by proton captures
instead of beta decay.  We note that the time scale considered
here is still sufficiently slow that no significant proton abundance
contributes to the r-process.  A study of this effect would require a time scale
at least of order 5 times  faster than the one adopted here.  
Such a study would also require
the implementation of many proton capture reactions for intermediate-mass
nuclei which is beyond the scope of the present network calculation.

\subsection{Neutron-Star Mass}
Nucleosynthesis in the r-process is strongly dependent on the
gravitational mass of the proto-neutron star (Wanajo et al. 2001).
Therefore, one can think of the neutron star mass as a parameter to be
adjusted to give good r-process yields.  A short expansion time is
required to obtain a large neutron-to-seed ratio at moderate entropy.
In our trajectory this expansion time is obtained by adopting a large
neutron-star gravitational mass ($M = 2.0 M_{\odot}$) and a neutron-star
radius of $10$ km.  Although, this mass is large compared with the
"standard" $1.4 M_{\odot}$ model,
an ideal condition for successful r-process could also have been
obtained with relatively rapid expansion time scale being preserved, 
for example, by altering the outer
boundary conditions in the hydrodynamic  model.  
Hence, one should not be too dismayed at this large neutron-star mass.

Furthermore, although a neutron-star mass of $2.0 M_{\odot}$ is large,
there are established equations of state (Shen et al. 1998, Weber 1999,
Sumiyoshi et al. 1995) which can stabilize neutron stars having masses
up to $M \le 2.2 M_{\odot}$.  This is also still consistent with
observed constraints on the maximum neutron star mass ranging $0.5
M_{\odot} \lesssim M \lesssim 2.0 M_{\odot}$ (Thorsett et al. 1993).
There is also other observational evidence supporting the existence of
massive neutron stars.  Massive supernovae are believed to have massive
iron cores $\geq 1.8 M_{\odot}$ and to leave massive remnants (Turatto
1998).  SN1994W and SN1997D are presumed to be due to $25 M_{\odot} - 40
M_{\odot}$ massive progenitors because the detected $^{56}$Ni abundance
is low (Sollerman et al. 1998, Turatto 1998).

Incidentally, a large dispersion in the heavy element abundances of halo
stars has been observed (McWilliams et al. 1995, Ryan et
al. 1996). Using an inhomogeneous galactic chemical-evolution model,
Ishimaru and Wanajo (1999) have shown that this observed dispersion
could be a natural consequence of r-process nucleosynthesis in
supernovae of massive $M \geq 30 M_{\odot}$ progenitors. Such progenitors
could conceivably have large core masses.

\subsection{r-Process Initial Conditions}

We start the r-process network calculation at a time when the
temperature has dropped to T$_9 = 9.0$.  We display the time variation
of the temperature (thin solid curve)
and mass density (dashed curve) in the top panel of Figure 2.
Time t $= 0$ s refers to the time at which T$_9 = 9.0$. From this point
the temperature drops very rapidly and then becomes almost constant at
around T$_9 \sim 0.62$.  The initial composition of the material is
taken to be free neutrons and protons with an electron fraction of Y$_e$
($= \rm{Y}_p$) = $0.42$. This was taken from the hydrodynamical
simulation of Sumiyoshi et al. (2000).

\section{Results}

Once equilibrium between (n, $\gamma$) and ($\gamma$, n) reactions
is obtained, the neutron-capture flow path runs through nuclei with
almost the same S$_n$-value along the nuclear chart.  The r-process
path strongly depends on what S$_n$-value is favored by the
neutrino-driven winds.  The optimal single-neutron separation energies,
S$_n$, assuming (n, $\gamma$) equilibrium, are given by
\begin{equation}
{\rm S}_n = \frac{{\rm T}_9}{5.040} \times \{34.075-\log({\rm Y}_n \times \rho)
+1.5 \times \log({\rm T}_9)\}~~[\rm{MeV}],
\label{saha}
\end{equation}
where $\rho$ is the total mass density of material in the expanding
neutrino-driven wind (in units of g~cm$^{-3}$), and Y$_n$ is the neutron
number fraction.  The calculated S$_n$ value along r-process path is
shown by the thick solid curve in the top panel of Figure 2.  We take the
time evolution of T$_9$ and $\rho$ from the wind model of Sumiyoshi et
al. (2000), and the Y$_n$-value is calculated using the present "full
network" code.  A typical S$_n$-value at freezeout is $2\sim4~{\rm MeV}$
in the literature.  However, in the present wind-model analysis, the
calculated S$_n$-value is $\sim 1$~MeV (Figure 2).  This is because the
expansion model has a short dynamic time scale and the material in the
neutrino-driven wind is very neutron rich.

Since T$_9$(t), $\rho$(t) and Y$_n$(t) depend on time, S$_n$ also varies
with time.  S$_n$ first decreases rapidly due to the expansion of the
wind while the $\alpha$-process operates at high temperatures
2~$\lesssim$~T$_9$.  The arrow at t = $3.3 \times 10^{-3}$ s indicates a
typical time at which the $\alpha + \alpha$ + n or triple $\alpha$
reaction produces seed nuclei. It eventually reaches S$_n \sim$ 1 MeV at
around t $\sim$ 20 ms, and stays almost constant until the r-process
freezes out, i.e.  when the time scale for neutron capture becomes
longer than that of the $\beta$-decay.  After the freezeout at t $\sim$
0.7 s, free neutrons are exhausted, and the nuclear reaction flow
$\beta$-decays towards the stability line.  Therefore, S$_n$ quickly
rises though the temperature remains almost constant.

We show the calculated seed abundance, Y$_S$, and the neutron-to-seed
abundance ratio, Y$_n/\rm{Y}_S$, as a function of time in the lower panel
of Figure 2.  Y$_S$ is defined as the sum of the number abundance
fractions of intermediate-to-heavy mass elements Y$_S = \bf\Sigma$ Y$_A$
($70 \le \rm{A} \le 120 $).  The solid and dashed curves respectively
display the results calculated in the "full network" and the "small
$\alpha$-network" codes.  We also show in Figure 3 the calculated final
abundance yields of the r-process elements for this trajectory.  
Also shown for comparison are the  relative solar
r-process abundances from K$\ddot{\rm a}$ppeler et al. (1989).

When we use the "full network" code, this  particular
trajectory happens to  more or less
reproduces the r-process abundance peaks near A $\sim$ 80, 130 and 195.
In the case of the "small $\alpha$-network" code, light-mass elements
with A $\lesssim$ 150 are underabundant.  The purpose of this
illustration, however, is not to argue that this is a better model for
the r-process.  Indeed, in most models the challenge has been to provide
enough neutrons per seed.  Here we see that the flow to heavier nuclei
is considerably diminished in the expanded network. Thus, for most
r-process models, this expanded reaction network, as necessary as it may
be, may actually make a bad situation worse.

This contrast between the two calculated results can be traced to
drastic changes in the seed production. As shown in the lower panel of
Figure 2, Y$_S$ is continuously supplied at 10 ms $\lesssim$ t in the
"full network" calculation. In the "small $\alpha$-network" calculation,
however, the seed production tends to diminish after 300 ms and
completely stops well before freezeout.  This suggests that a new
nuclear reaction flow to seed material must exist in the "full network"
which manifests itself at late times. This new channel sustains a high
abundance level of seed elements and dominates the reaction flow
throughout the r-process even up to the freezeout time.

\section{Reaction Flows}
Having identified that the production of seed material is quite a bit
different in the two network calculations it is important to now analyze
the critical reaction flows in detail as a fluid element expands through
the wind.  For this analysis we consider two times during the evolution.
One corresponding to the $\alpha$ process conditions early in the wind,
and one corresponding to the later $r$-process conditions.  These two
selected times are indicated by arrows and dots on the top panel of Figure
2.

Figure 4 shows the nuclear reaction flow at t = $3.3 \times 10^{-3}$ s.
>From the lower panel on Figure 2 one can  see that the seed abundance
$Y_S$ is just starting to form at this time in the wind.
 This  is the start of $\alpha$-process.
This is to be compared with  Figure 5 which shows the flow
at t = $0.567$ s  corresponding to near the end of the r-process
as identified on the bottom panel of Figure 2.  

In Figures 4 and 5, the relative abundances for Z $\leq 15$ in the N-Z
plane are shown by circles whose diameters are proportional to the
logarithm of the abundance yields Y$_A = \rm{X}_A/A$ as indicated.
Small dots denote the network range adapted in the present study: The
"full network" is used in the calculated results shown in the upper
panels (a) of Figures 4 and 5, and the "small $\alpha$-network" in the
lower panels (b).  The main reaction paths are indicated by arrows.  For
further clarification, the critical reaction flows to produce carbon
isotopes are shown in Figure 6.  Once formed, these carbon isotopes
quickly convert to heavier seed nuclei. Figure 7 shows the relative
abundances of neutrons $Y_n$, protons $Y_p$ and alpha particles
$Y_\alpha$.

\subsection{$\alpha$-Process}
>From Figure 2 we see that the start of the $\alpha$-process conditions
of Figure 4 (at t = $3.3 \times 10^{-3}$ s) corresponds to T$_9 = 3.4$,
and $\rho = 8.0 \times 10^4$ g cm$^{-3}$.  At this point the nuclear
statistical equilibrium is just shifting to produce a large abundance of
$\alpha$ particles, as evidenced on Figure 7.  At early times up to
nearly $10^{-2}$ s, the flow through the large and small networks are
quite similar.  In both Figures 4(a) and 4(b) that there are abundant
nuclei around the $\beta$-stability line even in our "full network"
calculations and the main path resides inside the network range of the
"small $\alpha$-network" for Z $\leq 15$ (see Figure 4(b)).

For $\alpha$-process conditions the main reaction flow is triggered by
the upper path in Figure 6, i.e. the $\alpha(\alpha
n,\gamma)^9\rm{Be}(\alpha,n)^{12}\rm{C}$ or $\alpha(\alpha
\alpha,\gamma)^{12}$C reaction.  The importance of this path was pointed
out by Woosley and Hoffman (1992) and Woosley et al. (1994).  Although
the side flows of
$\alpha(^3\rm{H},\gamma)^7\rm{Li}(n,\gamma)^8\rm{Li}(\alpha,n)^{11}\rm{B}$
and $^7\rm{Li}(\alpha,\gamma)^{11}\rm{B}$ also involve appreciable
nuclear reaction flow, the subsequent $^{11}\rm{B}(p,\alpha)^{8}\rm{Be}$
reaction returns this flow back into $\alpha$-particles as indicates by
the dashed arrows in Figure 6.  Thus, although there are plenty of
protons as well as neutrons and $\alpha$-particles present at this time
(cf. Figure 7), three-body and alpha-capture reactions of stable nuclei
are more efficient than neutron captures at this relatively high
temperature T$_9 = 3.4$.  This is generally the case in the early epoch
of the trajectory because both the temperature and density are still
high enough for these reactions to occur.  Therefore, we find very
similar reaction paths in the two different network ranges shown in
Figures 4(a) and 4(b).  A few new paths are evident in the full network,
e.g. $^{9}$Be$(n, \gamma$)$^{10}$Be$(\alpha, \gamma)^{14}$C, and
$^{14}$C$(n, \gamma)^{15}$C($\alpha,n)^{18}$O.  These, however, make
very little difference.

  We have included the possible three body two-neutron capture reactions
(Efros et al. 1996) such as $^{4}$He$(2n,\gamma)^{6}$He and
$^{6}$He$(2n,\gamma)^{8}$He, and so on as shown in Figure 1. No
significant flow was observed through this channel either at this time
or later in the evolution.  However, some two-neutron channels may
important in heavier nuclei as described below.

\subsection{$r$-Process}
 Once seed material has begun to assemble by $t \geq 0.01$ s (cf. Figure
2) some production of $r$-process nuclei begins.  As the temperature and
density of a fluid element diminish, charged-particle reactions become
progressively slower and eventually neutron capture becomes more
important. A classical r-process-like flow, i.e. (n,$\gamma$) and
$(\gamma$,n) reactions followed by beta decay, starts.

Differences in the reaction flow between the two networks become
apparent immediately.  One can identify two characteristic features of
the light element abundances on Figure 7.  One is that the alpha
abundance is almost the same, but is a little lower in the full network
calculation (by about 1 percent). Since alpha particles are the most
abundant nucleus, this small difference of $\delta$Y $\sim 10^{-3}$ has  
an influence on heavy element production.  The other is that the neutrons are
exhausted earlier in the full network calculation. The flow at t =
$0.567 $ s shown in Figures 5(a) and 5(b) corresponds to just before
freezeout at the end of the $r$-process when the material in the wind
has cooled to T$_9 = 0.62$, and the density decreased $\rho = 5.4 \times
10^2$g cm$^{-3}$.

The differences between the two network calculations can be traced to
the fact that in the full network the main path runs through very
neutron-rich nuclei on the N-Z plane (cf. Figure 5a).  The added flow
paths mean that both neutrons and alphas are more efficiently converted
to seed, as shown in Figure 7. This process is prohibited in the small
network (cf. Figure 5b).  Hence, their abundance along the trajectory is
lower.  The presence of more seed material means that neutrons are more
quickly exhausted in the wind.

In addition to the $\alpha$-induced reaction chains $\alpha(\alpha
n,\gamma)^9\rm{Be}(\alpha,n)^{12}\rm{C}$, there are two main flow paths
to form carbon seed present at this time.  They are almost equally
important.  These are the Be-isotope chain $\alpha(\alpha
n,\gamma)^9\rm{Be}(n,\gamma)^{10}\rm{Be}(\alpha,\gamma)^{14}\rm{C}$ and
the Li-B chain
$\alpha(\rm{t},\gamma)^7\rm{Li}(n,\gamma)^8\rm{Li}(\alpha,n)^{11}\rm{B}$
(see Figure 6). As for the Be-isotope chain, when the
$^{10}\rm{Be}(\alpha,\gamma)^{14}\rm{C}$ reaction is switched off, the
reaction flow changes to $^9\rm{Be}(n,\gamma)^{10}\rm{Be}(n,
\gamma)^{11}\rm{Be}(n, \gamma)^{12}\rm{Be}(\beta^-)^{12}\rm{B}$. We have
also studied what happens if the $^8\rm{Li}(\alpha,n)^{11}\rm{B}$
reaction is switched off. In this case, neutron capture on $^8\rm{Li}$
leads to $^9\rm{Li}$ which decays back to $^9\rm{Be}$.  It is to be
noted, however, that the results obtained by switching off either new
chain are similar to those calculated in the full network.  This is
because either the Be-isotope chain or the Li-B chain is still very
active even if the other chain is turned off. When both new chains are
turned off, however, the result is almost the same as that calculated in
the small network.  The presence of the two new chains in the full
network is therefore the main difference between two networks.


As a final remark we point out that we have also studied the
time-integrated nuclear-reaction flows.  This identifies the most
important main flow paths.  We carried out numerical calculations in
which the thermonuclear reaction rates times number abundances of
interacting nuclei were integrated from time zero to the freezeout time
of the r-process.  These quantities give the total intensity of the
nuclear reaction flow passing through each nucleus.  We find that the
main flow paths are almost the same as those indicated by the arrows in
Figure 5 which displays a snap shot at the time t = $0.567 $ s.  We thus
conclude that the main flow paths identified in Figure 5 (a) indicate the
significance of the new reaction channels for the production of the
final r-process abundance yields.

\subsection{Effects of Wind Time Scale and Neutrino Interaction}

  In a realistic supernova simulation one expects that the expansion
time in the wind will differ as the proto neutron star cools and the
bubble expands.  To identify the conditions at which the expanded full
network is important we have run simulations in the full and small
network for different expansion time scales.  The final abundances
calculated by using the full network (solid curve) and the small
$\alpha$-network (dashed curve) are summarized in Figure 8.  The
different time scales of $\tau_{dyn} = 5.1$, $53$, and $100$ ms
correspond to different trajectories obtained in the hydrodynamic
supernova model of Sumiyoshi et al (2000). Here we see that there is
little difference between the two networks when the expansion time scale
is slow.  For both networks, the seed build up is too efficient and
there is little production of elements for $A > 130$.

We have also studied the effects of neutrino processes (Meyer et
al. 1992, Meyer 1995, Fuller $\&$ Meyer 1995, Qian et al. 1997,
McLaughlin et al. 1996) in our full network calculations.
Among three wind models which have different time scales ($\tau_{dyn} =
5.1, 53,$ and $100$ ms), only the fastest wind with $\tau_{dyn} = 5.1$ ms 
leads to successful r-process nucleosynthesis.  The specific
neutrino-nucleon collision time scale is given (Qian et al. 1997) by
\begin{equation}
  \tau_{\nu} \sim 50 \times L_{\nu,51}^{-1} \left ( \frac {\langle
   E_{\nu}\rangle }{\rm MeV} \right ) \left ( \frac {r}{50 \, \rm km}
   \right )^{2} \left (\frac{10^{-41}\, \rm
   cm^{2}} {\langle\sigma_{\nu}\rangle } \right) \rm {ms},
\end{equation}
where $L_{\nu,51}$ is the neutrino luminosity for each species in units
of $10^{51}$ erg/s, $\langle E_{\nu} \rangle $ is the average neutrino
energy, r $ = 50$ km is the typical radius of the wind at which the
neutrino interactions occur most efficiently in the present model
calculations, and $\langle\sigma_{\nu}\rangle$ is the neutrino-nucleon
interaction cross section. Only the fastest wind satisfies $\tau_{dyn}
< \tau_{\nu}$, while the other two do not. Because of this fact,
the r-process for the wind model with $\tau_{dyn} = 5.1$ ms is not
hindered by the neutrino processes, especially the charged-current
interaction process $\nu_e + n \rightarrow p+e^-$. We can thus conclude
that  a short dynamic-time-scale model can lead to a successful
r-process, as has already been discussed in the  literature (Cardall $\&$
Fuller 1997, Qian $\&$ Woosley 1996, Otsuki et al.  1999, and Sumiyoshi
et al. 2000).

\subsection{Comment on Reaction Rates}
Although the previous  estimate of the cross
section for the three-body reaction $\alpha(\alpha
n,\gamma)^9\rm{Be}$ was presumed to only be good to one order of
magnitude (Kajino et al, 2000) due to the two unresolved channel
contributions from $^8\rm{Be}  + $ n and $^5\rm{He} + \alpha$, a recent
measurement (Utsunomiya et al. 2001) has confirmed that our adopted
cross section (Angulo et al. 1999) in the present calculations is
correct to within a factor of two. An exclusive measurement of the
$^8\rm{Li}(\alpha,n)^{11}\rm{B^*}$ cross sections has also been carried
out by Mizoi et al. (2000), and a large contribution from the processes
producing excited bound states of $^{11}$B$^*$ (Boyd et al. 1992, Gu et
al. 1995) have been confirmed.  For available nuclear reaction data,
reaction rates of (n,$\gamma$) for the other neutron-rich Z $\geq 5$
nuclei are generally larger than the ($\alpha$,n) and
($\alpha$,$\gamma$) rates, except for $^{18}\rm{C}(\alpha,n)^{21}\rm{O}$
and $^{24}\rm{O}(\alpha,n)^{27}\rm{Ne}$.

Abundance yields of extremely neutron-rich nuclei, $^{15}$B, $^{18}$C,
$^{23}$N, $^{24}$O, $^{27}$F, $^{28}$Ne, $^{35}$Na, $^{36}$Mg,
$^{41}$Al, $^{42}$Si, $^{43}$P, etc., exhibit the largest abundances
among isotopes for each atomic number Z in the neutron-rich environment
of Figure 5(a).  These nuclei, except for $^{18}$C and $^{36}$Mg, are on
or near the neutron-drip line (We know that $^{28}$Ne is not the most
neutron-rich nucleus at Z$ = 10$, however in this mass formula it is).
$^{17}$B is known to be stable against particle decay, and $^{19}$B also
is suggested to be stable theoretically, while $^{16,18}$B are unstable.
$^{29}$F and some heavier elements beyond the network range adopted in
the present studies are predicted to be particle bound in
some theoretical nuclear models. Quite recently, the neutron drip line
up to fluorine has been studied by the projectile fragmentation
experiments and $^{31}$F has proved to be particle stable, while
$^{24,25}$N, $^{25-28}$O (except for $^{26}$O) and $^{30}$F are unstable
(Sakurai et al. 1999).  Radiative two-neutron capture reactions, i.e.
$^{15}\rm{B}(2n,\gamma)^{17}\rm{B}(2n,\gamma)^{19}\rm{B}$,
may play a significant role in the r-process.  In the present
calculations we assumed that the (n,$\gamma$) and ($\gamma$,n) reactions
are in thermal equilibrium for the first neutron and then the second
neutron is subsequently captured. Although inclusion of the
(2n,$\gamma$) reactions did not change drastically the final result,
these rates are presumed to be lower limit. More elaborated theoretical
calculation of the (2n,$\gamma$) reaction has suggested that the
di-neutron correlation may increase these cross sections (Kamimura
2001).  Photodisintegration reactions of $^6$He and $^8$He and their
electromagnetic structure also have been studied experimentally (Aumann
et al. 1999; Iwata et al. 2000).  Extensive measurements of the nuclear
properties of heavier neutron-rich nuclei near the drip line, including
$^{17,19}$B and $^{29,31}$F, are yet to be carried out.

At T$_9 = 0.62$ and $\rho = 5.4 \times 10^2$g cm$^{-3}$ in Figure 5(a),
most of nuclei on the neutron-drip line satisfy Eq. (1) approximately.
However, for carbon and magnesium isotopes, two abundant nuclei $^{18}$C
and $^{36}$Mg are not on the neutron-drip line.  The neutron separation
energy of $^{19}$C is smaller than the value of Eq. (1) for sustaining
the steady-state flow at this temperature and neutron density.  As a
result, even at this low temperature, T$_9 = 0.62$, the
$^{18}\rm{C}(\alpha,n)^{21}\rm{O}$ reaction becomes faster than
$^{18}\rm{C}(n,\gamma)^{19}\rm{C}$.  Likewise,
$^{36}\rm{Mg}(\alpha,n)^{39}\rm{Al}$ is faster than
$^{36}\rm{Mg}(n,\gamma)^{37}\rm{Mg}$.  For this reason, the
(n,$\gamma$)A($e^- \nu$) r-process flow is broken before reaching the
neutron-drip line. This approximately satisfies the condition
represented by Eq. (1) at this time t = $0.567 $ s.  Precise
experimental studies of the reaction rates for the two competing
processes $^{18}\rm{C}(\alpha,n)^{21}\rm{O}$ vs.
$^{18}\rm{C}(n,\gamma)^{19}\rm{C}$ and
$^{36}\rm{Mg}(\alpha,n)^{29}\rm{Si}$ vs.
$^{36}\rm{Mg}(n,\gamma)^{37}\rm{O}$ would be highly desirable.

Recent progress at radioactive nuclear beam facilities has provided a
remarkable opportunity to study the nuclear structure and reactions of
extremely neutron-rich radioactive isotopes.  The neutron separation
energy of $^{19}\rm{C}$ along with its electric dipole distribution has
been measured by using the Coulomb dissociation method (Nakamura et
al. 1999). A precise determination of this quantity is important because
$ ^{18}$C is at a branching point between the $(n,\gamma)^{19}\rm{C}$ and
$(\alpha,n)^{21}\rm{O}$ reactions as discussed above. The same technique
was applied to find a large deformation of $^{32}$Mg (Motobayashi et al.
1995).  At t $= 3.3 \times 10^{-3}$ s (in Figure 4(a)), the main nuclear
reaction flow still stays near the stability line and does not reach
$^{32}$Mg. However, by the time t$= 0.567$ s (in Figure 5(a)), neutron
captures by magnesium isotopes form a side flow together with the
strongest flow along the sodium isotopes. (See the abundance of magnesium
isotopes). At some intermediate time between $3.3 \times 10^{-3}$ s and
$0.567$ s, this side flow plays an important role. Therefore, a phase
transition to large deformation at $^{32}$Mg may have a non-negligible
effect on the production of seed abundances.

A new magic number N = 16 has recently been found near the neutron drip
line (Ozawa et al. 2000), which may also affect strongly the seed
abundance distribution because $^{23}$N, $^{24}$O, and $^{25}$F are some of
the most abundant nuclei on the main flow path in Figure 5(a).  The
$^{17}\rm{O}(n,\alpha)^{14}\rm{C}$ reaction cross section has been
precisely measured from thermal to about 350~keV neutron energies
(Wagemans et al. 2001). This study indicates a cross section that is
lower by a factor of 2$\sim$3 than that of previous measurements.  All
of these new experimental studies, as well as future work, will be
important in order to clarify the role of light neutron-rich nuclei in
the high-entropy r-process.

Before closing this section, we should emphasize that our findings are
based on the r-process nucleosynthesis models in neutrino-driven winds
with a short dynamic time scale.  For example, if we calculate the
r-process abundance patterns using the trajectories with the relatively
long dynamic time scale adapted by Woosley et al. (1994), the final
patterns are almost the same for both the full and small networks. We
can understand the reason for this by comparing the dynamic time scale,
$\tau_{dyn}$, with the time scale for the alpha-capture process,
$\tau_{\alpha}$.  As the value of $\tau_{dyn}$ becomes larger than
$\tau_{\alpha}$, there is enough time to make seed nuclei by the
alpha-process. Thus, seed nuclei are mainly produced by
$\alpha$-captures, and the reactions of light neutron-rich nuclei are
unimportant.  In the alpha-process, the $\alpha(\alpha
n,\gamma)^9\rm{Be}(\alpha,n) ^{12}\rm{C}$ reaction is the key to make
heavy nuclei. Therefore, the value of $\tau_{\alpha}$ is regulated by
the $\alpha(\alpha n,\gamma)^9\rm{Be}$ reaction and is given (Meyer et
al. 1992) by
\begin{equation}
  \tau_{\alpha} \equiv \frac {1}{\rho ^2 {\rm Y}_{\alpha} ^2 {\rm Y}_n
  \lambda_{\alpha\alpha n} },
\end{equation}
where Y$_{\alpha}$ and Y$_n$ are the yields of alphas and neutrons, and
$\lambda_{\alpha\alpha n}$ is the rate of the $\alpha(\alpha
n,\gamma)^9\rm{Be}$ reaction. At a typical temperature for the
alpha-process, T$_9=5$, we have calculated the values of
$\tau_{dyn}$/$\tau_{\alpha}$. In the short dynamic time scale model,
$\tau_{dyn}$/$\tau_{\alpha} \sim 10^{-4}$, whereas
$\tau_{dyn}$/$\tau_{\alpha} \sim 10^{-1}$ for the model of Woosley et
al. (1994) (trajectory 40).  Thus, the alpha-process is relatively
efficient in the model of Woosley et al. (1994).

\section{Summary}

We have extended the network for Z $< 10$ nuclei including almost all
possible charged and neutron-capture reactions out to the neutron drip
line. We have studied the role of the reactions of light neutron-rich
nuclei in the r-process, utilizing a wind model of massive SNeII
explosions. We find that a new path for the r-process opens in these
neutron-rich nuclei from the "full network" calculations in models with
a short dynamic time scale. In our model calculation, the third
r-process peak at A$\sim 195$ is lower, while nuclei around A $\sim 50$
are more abundant than in the "small $\alpha$-network" calculations. 
This is because the available neutrons to make heavy nuclei
are diminished by neutron captures on the light seed nuclei from the
$\beta$-stability line to the neutron-drip line.  Note that these
results are restricted to models with a short dynamic time scale in
which the temperature drops rapidly and the alpha-process is relatively
unimportant. This result shows that light neutron-rich nuclei could be
important in other  very neutron-rich situations as well, 
such as a prompt explosion
(Sumiyoshi et al. 2001) or binary-neutron-star mergers.

Thus, we find that light neutron-rich nuclei can play an important role
in r-process nucleosynthesis in models with a short dynamic time
scale. The yields of even the most neutron-rich isotopes can be abundant
along the r-process path.

\acknowledgments

One of the authors (MT) wishes to acknowledge the
fellowship of RIKEN Junior Research Associate.  One of the authors (GJM)
also wishes to acknowledge the hospitality of the National Astronomical
Observatory of Japan where much of this work was done.  This work has
been supported in part by the Grant-in-Aid for Scientific Research
(10640236, 10044103, 11127220, 12047233) of the Ministry of Education,
Science, Sports, and Culture of Japan, and also in part by DoE Nuclear
Theory Grant (DE-FG02-95-ER40394 at UND).  The numerical simulations
have been performed on the supercomputers at RIKEN.

\newpage

\begin{table}[h]
\begin{center}
\begin{tabular}{c|c|c|c}\hline \hline
  Z  & A$_{min}$ & A$_{max}(\rm{small}~\alpha\rm{-network})$ &
A$_{max}(\rm{full~network})$\\ \hline \hline
 H   & 1    & 3   & 3 \\
 He & 3    & 4   & 4 \\
 Li  & 6    & 7   & 9 \\
 Be & 7    & 9   & 12 \\
 B   & 8    & 11 & 15 \\
 C   & 11  & 14 & 20 \\
 N   & 13  & 15& 23 \\
 O   & 15  & 18 & 24 \\
 F   & 18  & 20 & 27 \\  \hline
 Ne & 19  & 28 & 28 \\
 Na & 22  & 35 & 35 \\
 Mg & 23  & 38 & 38 \\
 Al  & 26  & 41 & 41 \\
 Si  & 27  & 42 & 42 \\
 $\cdot$  & $\cdot$ & $\cdot$ & $\cdot$ \\
 $\cdot$  & $\cdot$ & $\cdot$ & $\cdot$ \\
 $\cdot$  & $\cdot$ & $\cdot$ & $\cdot$ \\
 Am   & 241   &  279  &  279\\  \hline \hline
\end{tabular}
\end{center}
\vspace{0.5cm}
\caption{ Comparison of the "small $\alpha$-network" and the "full
network" for light-mass nuclear systems.  The nuclides 
(A$_{\rm{min}}\leq \rm{A} \leq \rm{A}_{\rm{max}}$) included in the
networks are different from each other only for Z $ < 10$.}
\end{table}

\newpage

\begin{figure}[t]
\epsscale{1.0}
\plotone{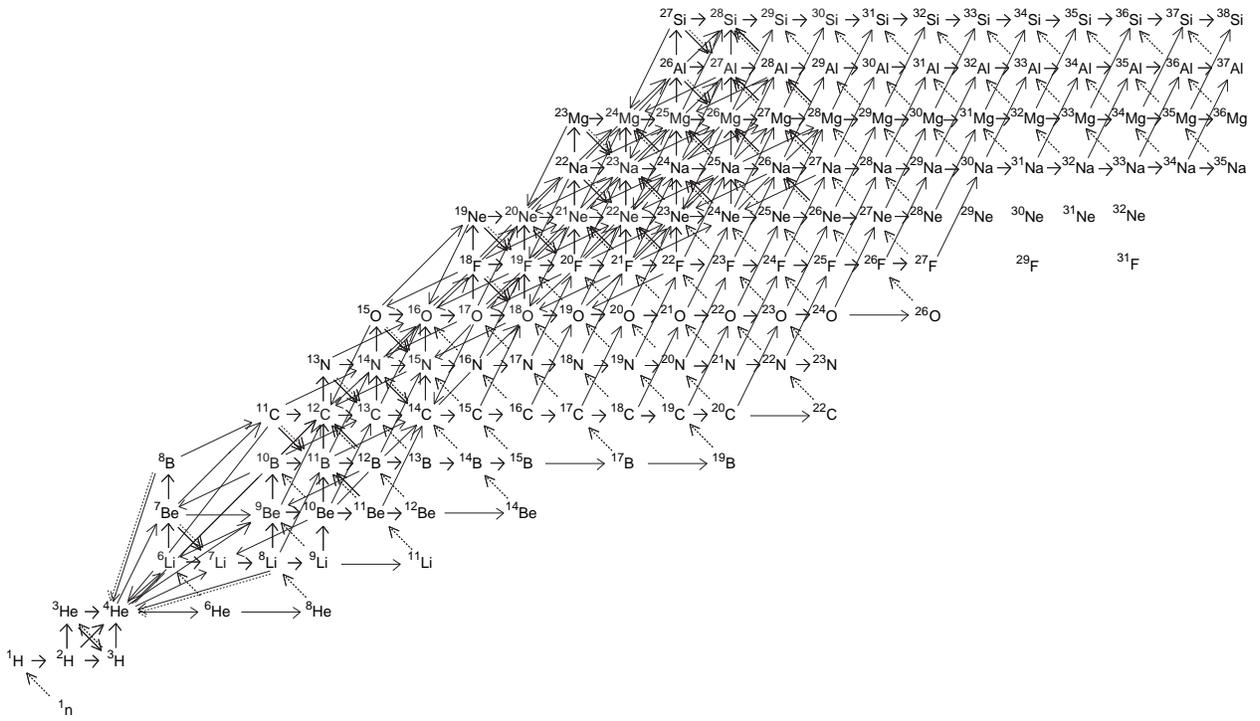}
\caption{Light-mass region of the extended nuclear reaction network
("full network")
used for the present r-process nucleosynthesis calculations.}
\label{network}
\end{figure}

\begin{figure}[t]
\epsscale{1.0}
\plotone{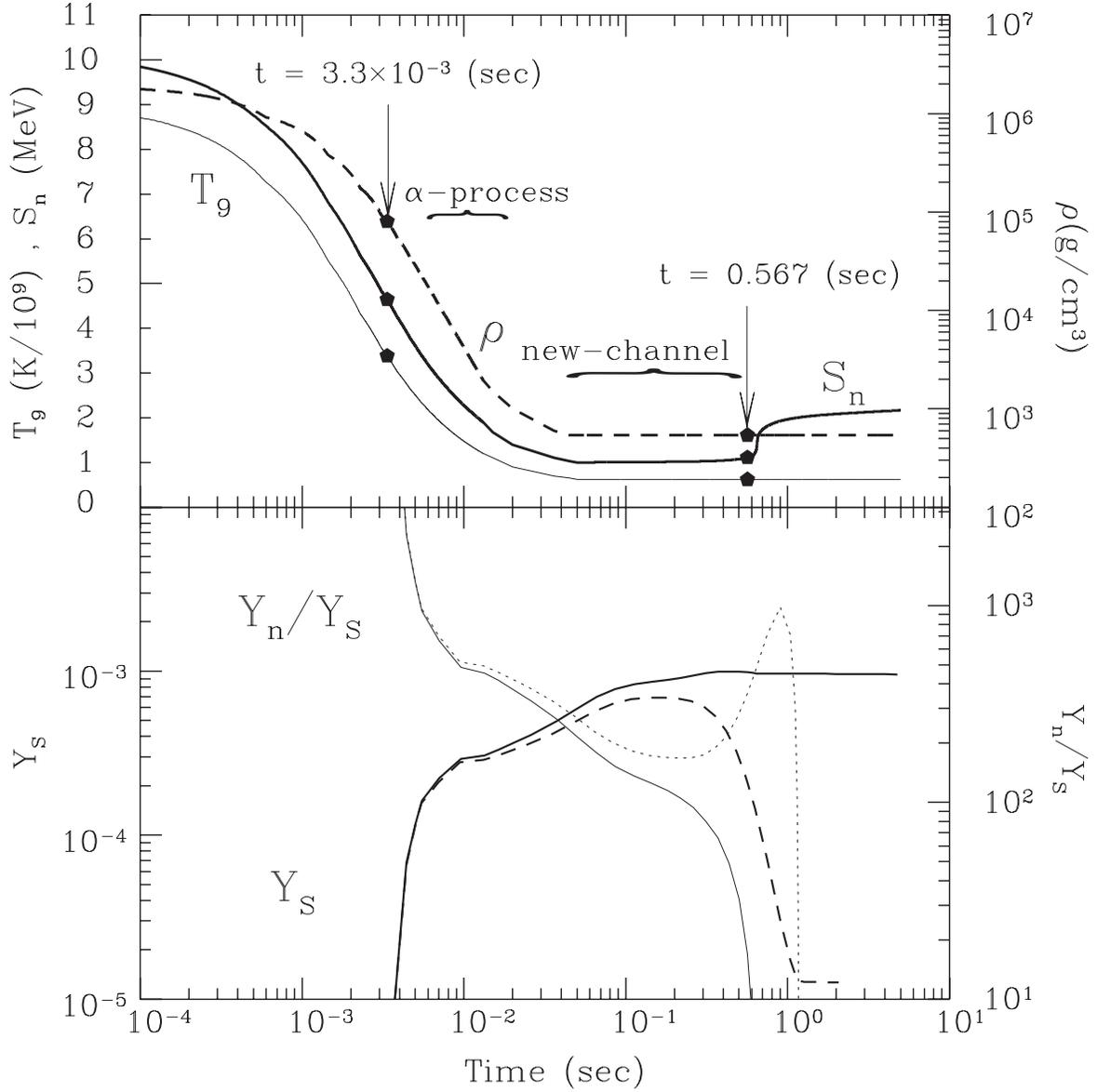}
\caption{The upper panel shows the temperature T$_9$ (thin solid curve)
and mass density $\rho$ (dashed curve) as a function of time for the
neutrino-driven wind model adopted in the present numerical simulation
(Sumiyoshi et al. 2000).  The thick solid curve is the neutron
separation energy, S$_n$, assuming steady-state flow. The arrow at t $=
3.3 \times 10^{-3}$ s indicates a typical time at which the $\alpha
\alpha$n or triple $\alpha$ reactions predominantly lead to the
production of seed nuclei (70 $\le$ A $\le$ 120).  The arrow at t =
$0.567$ s indicates a typical time at which the light-mass neutron-rich
nuclei contribute significantly to the production of seed nuclei. In the
lower panel, the thick solid and thick dashed curves display the seed
abundance Y$_S$ calculated in the "full network" (solid) and in the
"small $\alpha$-network" (dashed) respectively.  Also shown by the thin
solid and thin dashed curves are the neutron-to-seed ratios,
Y$_n$/Y$_S$, for the "full network" (solid) and the "small
$\alpha$-network" (dashed).
\label{Sn}}
\end{figure}

\begin{figure}[t]
\epsscale{1.0}
\plotone{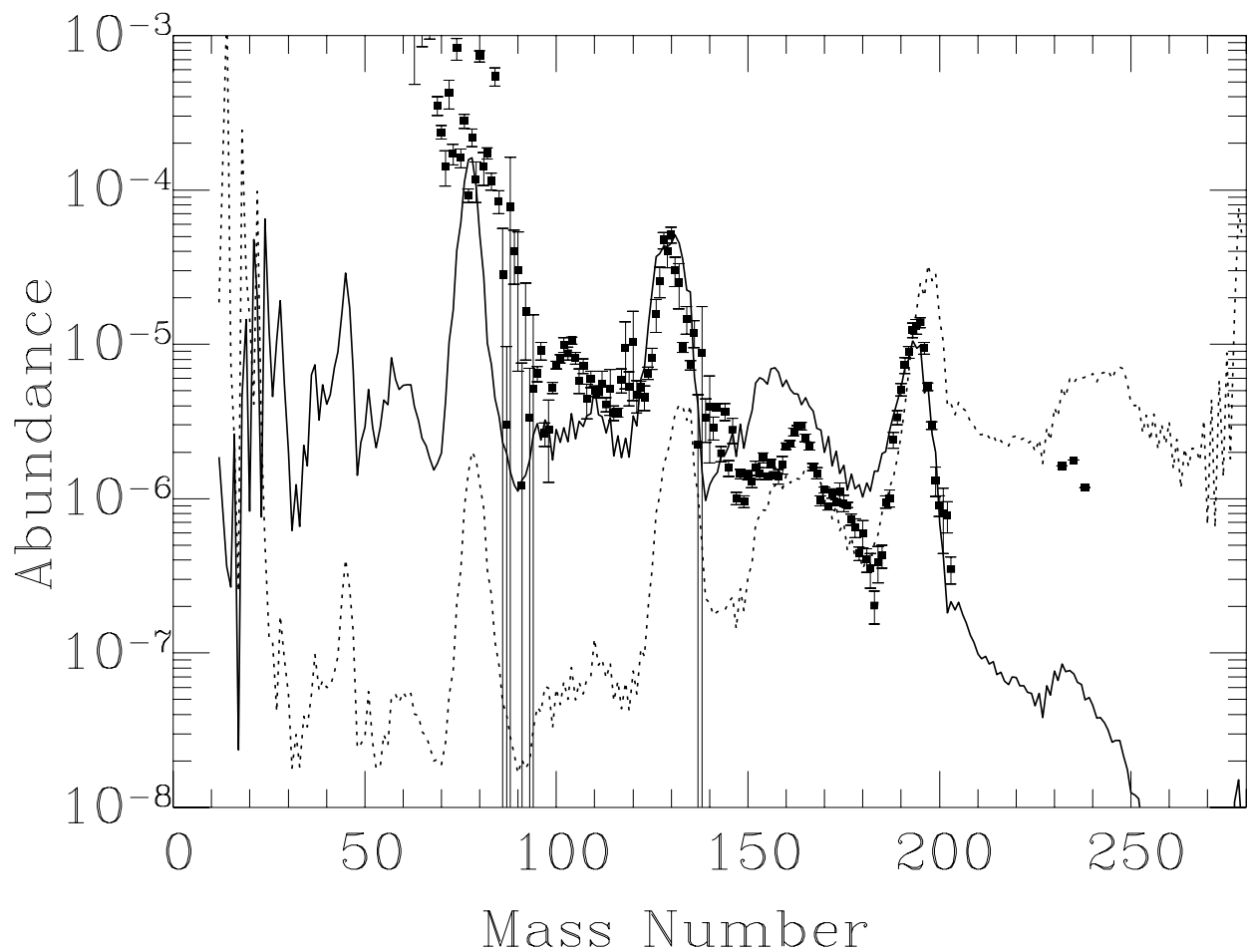}
\caption{Final abundances as a function of mass number in the present
r-process calculations.  The solid line is the calculated result
obtained by using the "full network", and the dotted line is from the
"small $\alpha$-network".  Closed squares are the data points of the
solar-system r-process abundances (in arbitrary unit) from
K$\ddot{\rm a}$ppeler et al. (1989).
\label{final}}
\end{figure}

\begin{figure}[t]
\epsscale{1.0}
\plotone{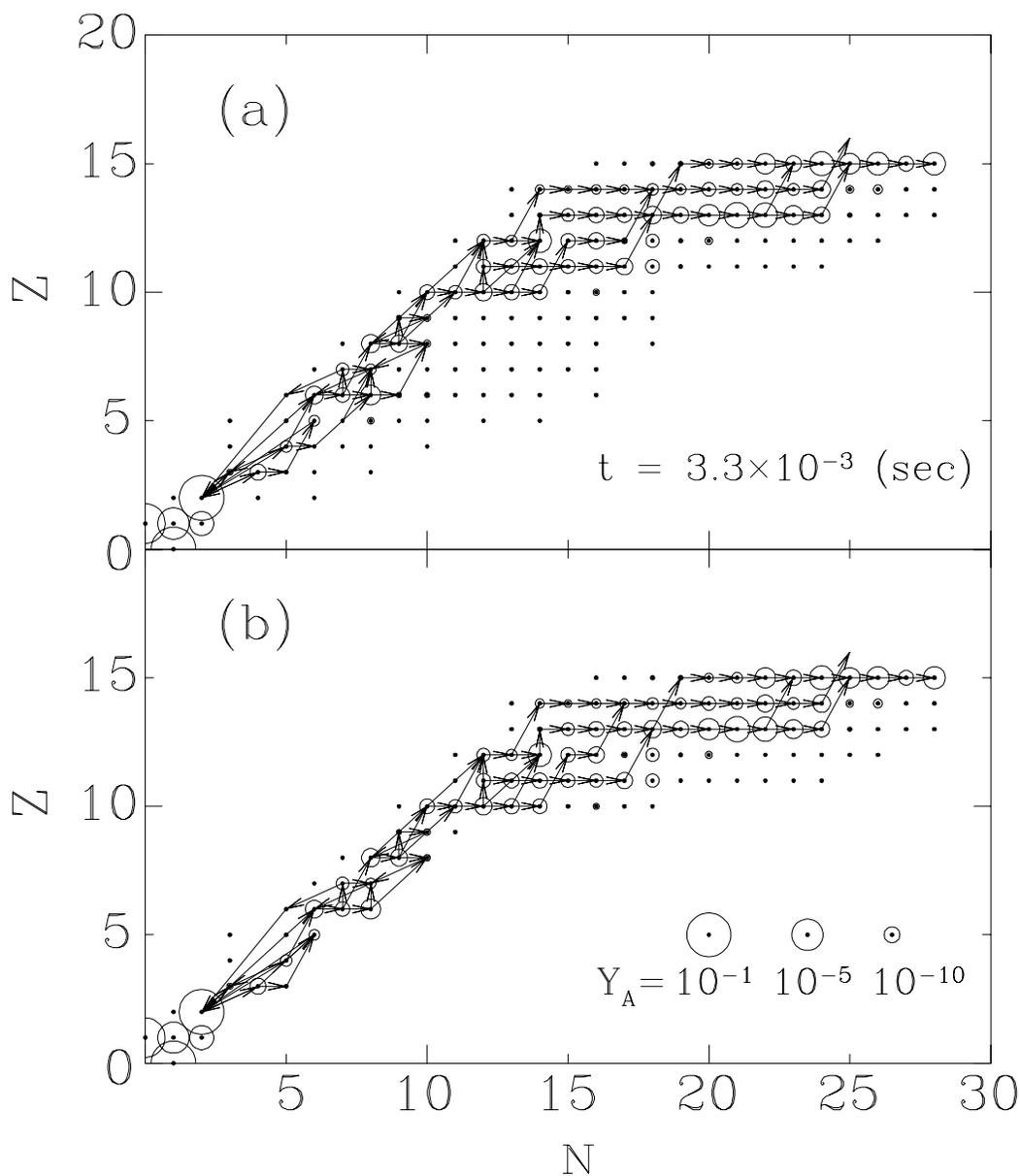}
\caption{Calculated abundances Y$_A$ of the light-to-intermediate mass
elements for (a) the "full network" and (b) the "small
$\alpha$-network", at t$ = 3.3 \times 10^{-3}$ s and relatively high
temperature T$_9$ = 3.4, and density $\rho = 8.0 \times 10^4$ g
cm$^{-3}$, as indicated by the left arrow in Figure 1.  Arrows in this
figure indicate the main nuclear-reaction flow paths.
\label{abundance.early}}
\end{figure}

\begin{figure}[t]
\epsscale{1.0}
\plotone{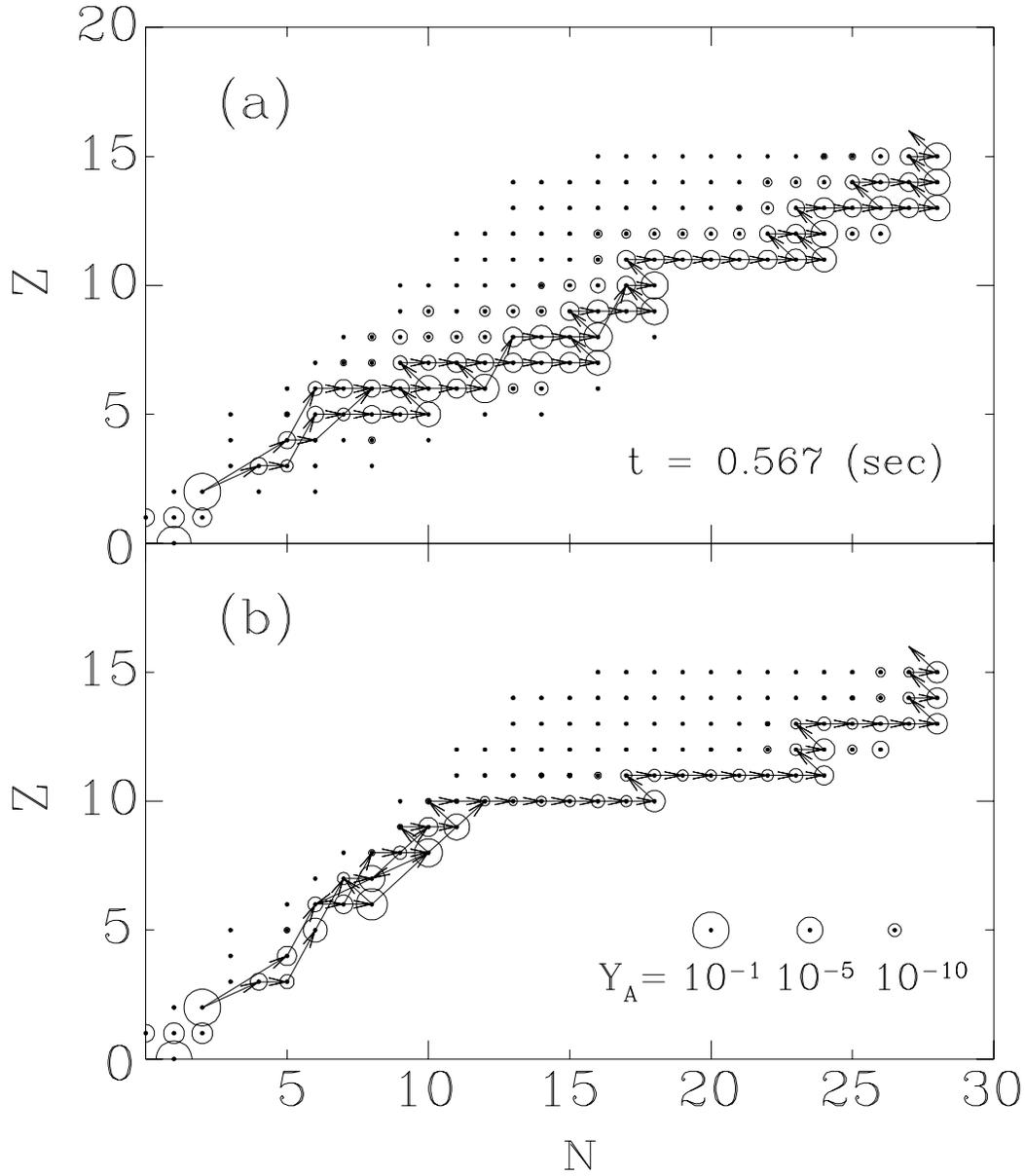}
\caption{Same as Figure 4 for (a) the "full network" and (b) the "small
$\alpha$-network", but at t=$0.57 $ s, T$_9 = 0.62$, and $\rho = 5.4
\times 10^2$ g cm$^{-3}$, as indicated by the right arrow in Figure 1.
\label{abundance.late}}
\end{figure}

\begin{figure}[t]
\epsscale{1.2}
\plotone{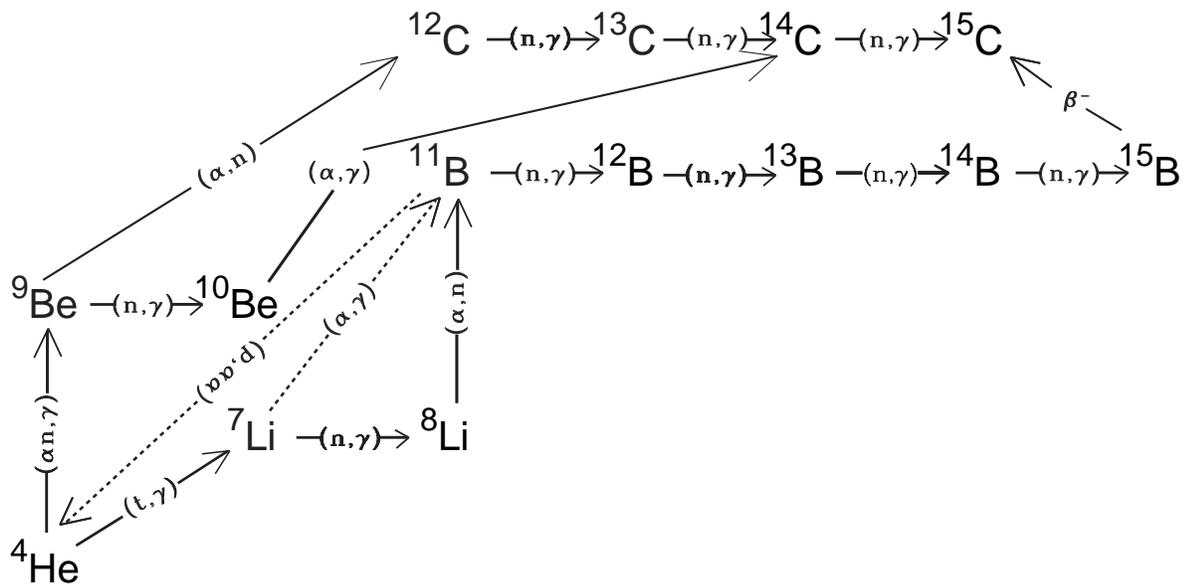}
\caption{Critical reaction flows to produce carbon isotopes. For
$\alpha$-process conditions at earlier times the main reaction flow is
triggered by the upper-most path with the side flow as indicated by the
dashed arrows. For r-process conditions at later times three reaction
paths shown by solid arrows are almost equally important. See the text
for more details.
\label{chart.kakudai}}
\end{figure}

\begin{figure}[t]
\epsscale{1.0}
\plotone{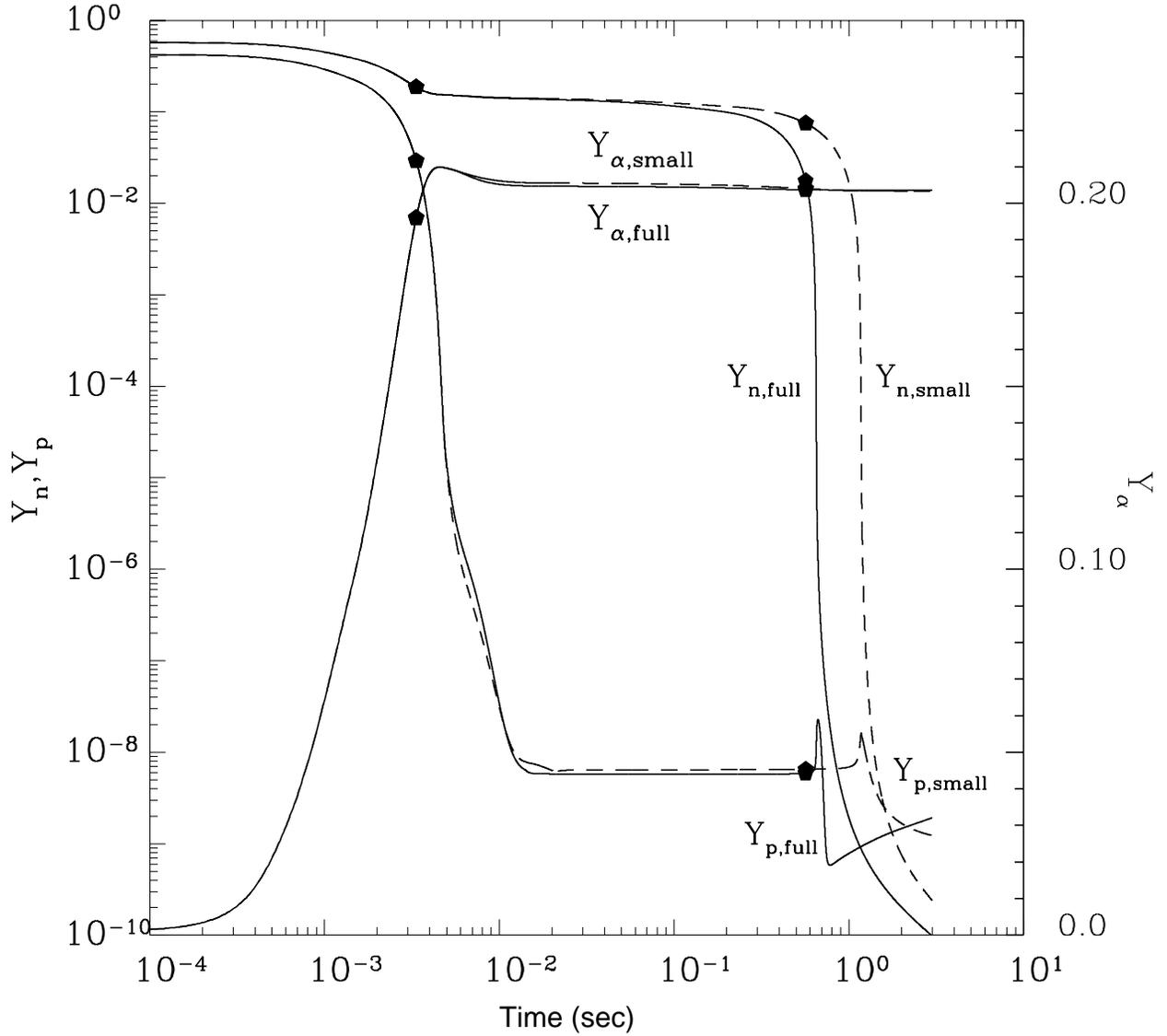}
\caption{Neutron, proton, and $\alpha$-particle number fractions as a
function of time. The solid and dashed curves are the number fractions
calculated in the "full network" (solid) and in the "small
$\alpha$-network" (dashed), respectively. Closed dots indicate typical
times t$ = 3.3 \times 10^{-3}$ s for the $\alpha$-process and t=$0.57 $
s for the r-process as in Figure 2.
\label{abundance.npa}}
\end{figure}

\begin{figure}[t]
\epsscale{0.9}
\plotone{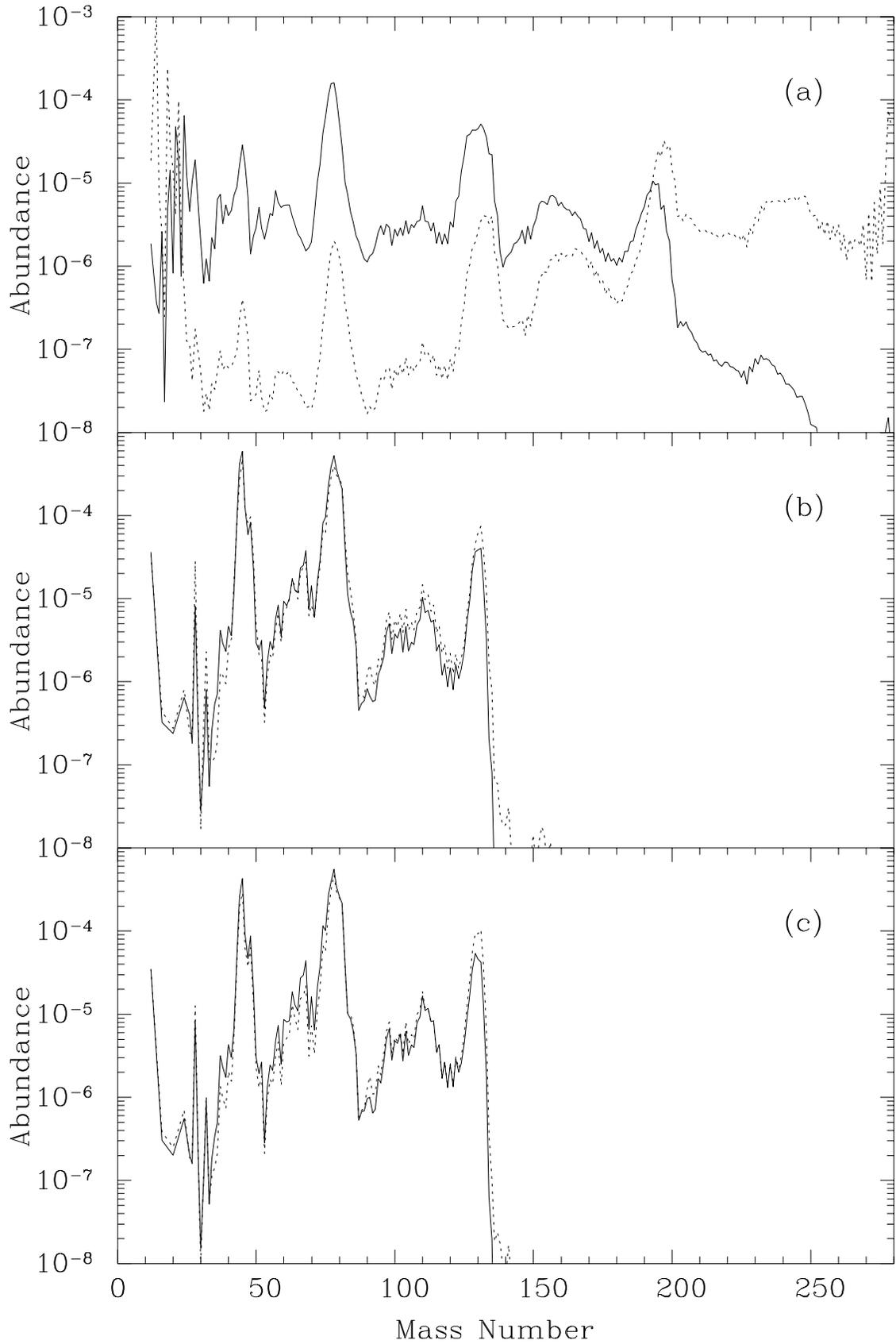}
\caption{Final abundances as a function of mass number, calculated by
using the "full network" (solid) and the "small $\alpha$-network"
(dashed) for three different dynamic time scales (a)$\tau _{dyn} = 5.1$
ms, (b)$53$ ms, and (c)$100$ ms.
\label{abundance.time}}
\end{figure}

\end{document}